\documentclass[conference]{IEEEtran}
\usepackage{cite}
\usepackage{url}
\usepackage{amsmath,amssymb,amsfonts}
\usepackage{algorithmic}
\usepackage{graphicx}
\usepackage{csquotes}
\usepackage{textcomp}
\usepackage[table]{xcolor}
\usepackage[most]{tcolorbox}
\usepackage{colortbl}
\usepackage{array,tabularx}
\definecolor{tablerow1}{RGB}{230,230,230}
\definecolor{tablerow2}{RGB}{245,245,245}
\usepackage{ragged2e}

\newtcolorbox{promptbox}{
  colback=gray!10,
  colframe=gray!50,
  boxrule=0.5pt,
  arc=5mm,
  boxsep=5pt,
  left=5pt,
  right=5pt,
  top=5pt,
  bottom=5pt,
  fontupper=\ttfamily
}

\begin{document}

\title{\Large\bf Building Living Software Systems with Generative \& Agentic AI}

\author{
\IEEEauthorblockN{Jules White}
\IEEEauthorblockA{\textit{Department of Computer Science} \\
\textit{Vanderbilt University} Nashville, TN, USA \\
jules.white@vanderbilt.edu}
}

\maketitle

\begin{abstract}
This paper is an opinion paper that looks at the future of computing in the age of Generative \& Agentic AI. Current software systems are static and inflexible, leading to significant challenges in translating human goals into computational actions. "Living software systems" powered by generative AI offer a solution to this fundamental problem in computing. Traditional software development involves multiple layers of imperfect translation, from business requirements to code, resulting in rigid systems that struggle to adapt to changing user needs and contexts. Generative AI, particularly large language models, can serve as a universal translator between human intent and computer operations. This approach enables the creation of more flexible, context-aware systems that can dynamically evolve to meet user goals. Two pathways for implementing living software systems are explored: using generative AI to accelerate traditional software development, and leveraging agentic AI to create truly adaptive systems. New skills like Prompt Engineering are necessary. By reimagining software as a living, adaptable entity, we can create computing interfaces that are more intuitive, powerful, and responsive to human needs.
\end{abstract}

\begin{IEEEkeywords}
Large language models (LLMs), Prompt engineering, Agentic AI, Agents, Generative AI, Living software systems
\end{IEEEkeywords}

\section{Static is Dead: Living Computing Systems are the Answer}

We build software that is static and dead. Alan Watts, the author of the \textit{Wisdom of Insecurity}~\cite{watts1962wisdom}, aptly summaries the problems of things that are static when he states "[t]he more a thing tends to be permanent, the more it tends to be lifeless." Most of our software today is dead and lifeless. The world is constantly moving and changing and our interface to computing, called software, is static. We can change it, but at enormous time and cost. Our way of controlling software is limited and dead.

We pretend that this isn't the case. We budget for software as if it is a one-time expense. We refuse to hit "update" when it pops up on our computer out of fear that something will break. We don't budget for software as a continually evolving translation problem. We don't view software as a race to perpetually keep up with the "goals" of the people in the world and make sure the software can accurately translate them into computing. 

So, what choice do we have? We work with inflexible languages, abstractions, and systems in computing. We have hordes of tools that don't talk to each other and break when one tools decides to change slightly. How do we ever extract ourselves from this terrible mess?

\textbf{The answer is "living software systems" that are built from the beginning to translate and adapt.} Systems that are living and that don't suffer from the "permanence" that Watts' points out makes things "lifeless". Amazingly, this type of "living computing system" is within our grasp. 

\subsection{Enter Generative AI}

The Transformer Model, the deep learning architecture behind Generative AI, was built for translation~\cite{vaswani2017attention}. It was built and tested on language translation tasks. This fundamental purpose of translation is at the core of how generative AI will revolutionize our interaction with computers. Just as the Transformer Model bridges the gap between different human languages, it will bridge the gap between human intent and computer operations. For instance, instead of manually inputting each expense into a system, categorizing it, and attaching receipts, users will simply tell the AI, "Submit my travel expenses for the Chicago conference last week," and the AI will handle the entire process.

The interface to Generative AI is "prompts." What are prompts? Prompts are simply human language. In many cases, these are human language that expresses a goal to be solved. Let's look at some examples:

\begin{displayquote}
"Write an email to my email client telling it that I quit! I am done putting up with this terrible abstraction for human communication."
\end{displayquote}

Does this prompt express a goal? Yes. Could any commercial software system before ChatGPT interpret this goal and help you with it? No. Can ChaGPT, Claude, LLama, and Gemini? Absolutely. Can they translate the goal into text? Yes.

What about if we change the prompt up?

\begin{displayquote}
"Yo! I am sick of email. Tell my email client I am breaking up with it."
\end{displayquote}

Does it matter that I changed how I expressed my goal? No. Generative AI can accurately translate it into text. However, unlike our current software, it can also pick up in the nuance of expression. 

The next iteration of Generative AI, called Agentic AI, not only can translate goals into text, but it can take actions on our behalf with software tools, or the Application Programming Interfaces (APIs) that sit beneath them. For example, the following prompt will teach GPT-4o to start translating requests into a set of sample actions for an email client:

\begin{displayquote}
My email has the following actions, compose:[insert message], which will compose a message and put it in my drafts, read, which will list the most recent draft, and send, which will send the most recent draft. Whenever I tell you to do something, decompose it into one or more of these actions.

Always produce your output in the format:

Thought: [Insert your thinking in one sentence]

Action: [Insert Action]

Action Details (optional): [Insert any details]
\end{displayquote}

Now, if you follow up with a goal, like "write an email to my boss asking for a raise", it will translate it into a message that has one or more actions that it is going to take to accomplish the goal:

\begin{displayquote}
Thought: To write an email to your boss asking for a raise, I need to compose a message and put it in your drafts.

Action: Compose

Action Details (optional):

Subject: Request for Salary Review

Message:...it's email to your boss..
\end{displayquote}

It is a short hop from this translation to a message with a series of "actions" to actually running those actions in software. In fact, OpenAI's Custom GPTs already possess this capability~\cite{OpenAI2023GPTs}. A Custom GPT can be configured to know about APIs and automatically talk to them to accomplish goals. Building an expense tracking app is as simple as giving a Custom GPT access to actions to get the rows in a spreadsheet and add a row to a spreadsheet. Want it to draft a reimbursement email to? Just give it access to a "send email" action and an associated API. Want it to enforce a travel policy? Just tell it the rules. 

The word "translation" aptly captures the role that generative AI will play in computing. It will serve as a universal translator, converting user intentions expressed in natural language into the specific "languages" of various software tools and APIs. In our travel expense example, the AI will automatically scan receipts from the user's email or phone, categorize expenses, convert currencies if needed, apply the company's reimbursement policies, and submit the report through the appropriate channels.
This transformation will dramatically simplify how we interact with computers. Rather than taking years to build a static software application, it will take hours and consist of giving an Agentic AI system access to the right APIs and a list of the correct human language instructions.

What happens if the travel policy changes? We just tell the Agentic AI that the policy has changed. We decide to add a new feature to handle expenses for sidewalk scooters because our employees are demanding it -- just tell the Agentic AI. We dramatically speed up the development cycle to update the software and make our systems able to adapt as the world changes.

Instead of learning multiple interfaces and manually bridging gaps between different applications, users will express their goals in natural language. The AI will then handle the translation into the appropriate series of actions across various software tools. This will lead to more intuitive computing experiences, reduce the learning curve for complex tasks, and democratize access to powerful computing tools for a wider range of users. In the case of travel expenses, what once required knowledge of specific expense categories, company policies, and system interfaces will become as simple as saying, "Process my business trip expenses." The AI will handle all the complexities, from currency conversion to policy compliance, making the process accessible to everyone in the organization regardless of their familiarity with the financial systems.

\section{Computing is a Translation Train Wreck}
The current world of computing is a translation train wreck. Every step of the way in computing, we are translating one thing to another, a goal to a choice in software tool, a business requirement into code, a real-world concept into a data structure. Except, our translations are always compromised, inaccurate, and static. 

What is translation? The Merriam Webster definition of translation includes "a rendering from one language into another" and "a change to a different substance, form, or appearance: CONVERSION." We fail to see how much translation is around us at every moment, particularly in the world of computing. 

The biggest transformation that is completely missed in all of this is that Gen AI will reshape computing and bring life back into it. Why? Software is a "translation" tool. Software translates what the user wants to do into a series of computer API calls to accomplish a task. All of the individual pieces of software are mini translators that speak some narrow language and mostly refuse to talk to each other. Users are forced to translate something they want to accomplish into the interfaces of 2-3 different tools, with completely different menus, buttons, etc. For example, submitting a travel expense reimbursement often requires juggling between an expense tracking app, a company's financial system, and possibly a separate approval workflow tool.

Let's think about all the forms of "interfaces" to software. The travel expense might start its journey when we receive an email about a flight booking. We have an email interface with rows of messages, the ability to send, receive, and forward messages. We can recategorize messages. Now, we need to translate "track" this expense so I can reimburse it later into a series of actions in this interface. You might forward the email to yourself and change the subject to make it easier to find later. You might move it to separate folder. You might plan on doing nothing and using search to find it later. The key point is that you have to translate this simple goal "track my expense" into the primitive capabilities of the tool and its actions -- which may not have been designed to do what you want to do.

If you are lucky, you have a tool that was exactly designed to achieve your specific goal. Except, the tool's interface is designed based on how someone else thinks you should track expenses or how their usability expert thinks you should or maybe how their business model thinks you should. The interface isn't designed for YOU. Yet again, you are stuck translating.

Now, you decide you want to compile all of your receipts and turn them into an expense report. You may be using a completely different tool again. Perhaps you begin entering the expense into the rows and columns of a spreadsheet. You are using another interface.

Why didn't the spreadsheet just directly talk to the email program to get the expense? Because they don't speak the same language, understand the concepts (e.g., email thread vs. cell), and mostly refuse to help each other. We hear about "interoperability", but it is limited or doesn't exist in most cases.

\section{Software Engineering is a Sequence of Translation Mistakes}

Building software, which we call "software engineering", is rife with translation errors. We call it "engineering", as if we know what we are doing. We pretend that we have a structured process for "engineering" software. Instead, we tend to have a lot of smart people working together on a team performing ad-hoc translation of millions of details to create a tool that helps a user accomplish one or more of the goals we actually care about. 

Let's consider how the concept of "translation" permeates every level of software engineering. Just as with the travel expense example, software engineers are constantly engaged in acts of translation, often without realizing it.
Take the process of turning a business requirement into working code. It starts with a product manager saying something like, "We need a feature that allows users to split their expenses with friends." Sounds simple. We write these goals down and call them "requirements" so that we can track how well the system helps achieve them.

After writing our code, we need to translate it into something the computer can execute. Compilers and interpreters translate our human-readable code into machine instructions. And don't forget the database - we need to translate our object-oriented model of expenses and friends into a relational schema for efficient querying and updating.

Throughout this process, we're constantly losing information and nuance. The rich, context-heavy world of human experience gets boiled down to bits and bytes, SQL queries and API calls. It's like translating a poem from one language to another - you capture the general meaning, but subtleties and cultural context often get lost.
Here's the kicker - once we've gone through all these layers of translation, we then expect users to translate their needs back into the language of our software. We give them buttons to click, text fields to fill out, dropdown menus to select from. But none of these map directly onto the user's mental model of "splitting expenses with friends." They're forced to learn our language, our way of thinking about the problem.

\section{Current Software Lacks Context}

The problem with how we think about "goals" is that we pretend that they are independent entities that can be reasoned about without context. Can you really design the "best" interface for tracking expenses for every moment in my life? No. I may be perfectly capable of taking a picture of a receipt with my phone sometimes. Other times, I may be rushing in the airport with coffee in my hand. Or, I may have just received that expense via email or text message. No single interface is going to work.

The other problem is that our interfaces to computing limit the ability for a user to provide "context". There is no "context" of "Bob texted me a picture of the receipt for dinner last night that we need to split and I am driving to the airport" when I need to track that expense. There is just a "add expense" button with a complicated menu system to split it. 

Can you really translate a goal into actions to accomplish the goal without context? Goal: "I need to get to the other side." Of what?! Without context, there is no way to translate the goal into actions. Despite this important consideration, our current software systems have little to no ability to incorporate context into them. The user is just supposed to figure it out. Rather than simplifying the problem, so that the user can solve bigger and harder problems, we add layers of complex translation into the software systems that control our computing.

\subsection{We Train Users to Translate to the Interface, Rather than Adapting the Interface to the User}

Now, you say "Sure! We can support all of those requirements". We will build an interface that helps enter receipts while driving to the airport and supports text messages with receipts from people named Bob. Except, you build a software application that is so overly complicated and filled with menus and buttons that I can't figure out how to actually track an expense. It does everything you asked, except not in a way that is actually usable to me. If I can't figure out how to translate my original goal into the actions in the tool, the tool is dead and useless to me.

Of course, now we just say that we have "a training problem." Roll out the vendor training for everyone. If you can't figure out how to use the tool, it's because your users haven't been trained! Now, we will teach everyone to follow the superior thinking of the handful of software engineers, business analysis, project managers, executives, and other "stakeholders" that pretend that they chose the design. In reality, they made thousands of compromises to the design along the way just to get it to work.

Finally, we have all of our users trained, so what should we do? We build the next version that can solve even more problems. The first thing we do, in addition to packing even more incomprehensible menu icons onto the screen, is completely redesign the user interface. Had we known about all the new features, we would have designed it differently in the first place. Let's fix all of this and change the interface.

Guess what? Now, many of our users have no idea how to translate goals into the tool anymore. Now, we need to build more training and teach everyone all of the new stuff. All of this is so that they can translate their goals into the capabilities of the tool. 

\subsection{Technical Debt is the Accumulation of Old Translation Errors}

Yes, of course, along the way, we realized that we translated the requirements into an imperfect design. We call these hindsight translation errors "technical debt". We promise to go and fix these translation errors in the future. We never do, so the software engineers that follow us face a much more daunting translation problem of translating requirements into a language that someone else designed that can't accurately capture the concepts in the domain. 

We have barely scratched the surface of the translation problems in software. Translation permeates every level of software, from the button clicks all the way down to the bytes. Even team dynamics suffer from "specification ambiguity", which means that two different engineers translate a phrase in the plans into different concepts in their heads.

\section{The Single Most Important Generative AI Use Case}

Everyone is still talking about "Return on Investment" and "finding the right use cases" for Generative AI. The use case is obvious. We need to fix our translation problem in computing and software. We need to use this new technology to build living software systems that can dynamically evolve to accomplish a goal based on the actual context or to evolve our current software approaches much faster.

There are two distinct ways of using Generative AI to building living software systems: 1) we use Agentic AI; and 2) we use Generative AI to translate requirements into source code. The second example is already happening at a very rapid pace. Estimates put Generative AI-written code at 40\% of all code checked into GitHub~\cite{Microsoft2023}. 

The second pathway, using Generative AI to write systems that help us build software is the immediate impact of this technology. We simply teach humans to pair with the technology and use it to build much faster. The major problem that I see is that current system design approaches, languages, and platforms were designed by humans for humans. We need to begin building languages, architectures, documentation, and platforms that are focused on Generative AI first. Either way, these tools will allow us to build traditional static software much faster.

However, we need to realize that the conversations that we are having with Generative AI to produce the software are more important than the software itself. The human requirements and goals expressed in the developer's conversation are more important than the actual code. We can always translate these goals and requirements into code again from the conversation. The conversation is the intellectual property and not the code. Our tools don't account for this yet. 

The first pathway, using Agentic AI to build truly living systems, is the more radical approach, but it is the future. In this approach, the Generative AI sees the various APIs that underlie our traditional software as tools it can use to accomplish a goal. It dynamically translates user goals into actions with these tools. 

Programs for these tools are hidden prompts, often called "system prompts" or "root prompts", that sit behind the scenes providing the guardrails and strategies it uses to solve problems. An entire program might consist of:

\begin{displayquote}
"Any time I give you a travel expense, check my existing list of travel expenses using the Get Expenses API, make sure it isn't a duplicate, and then enter it with the Add Expense API." 
\end{displayquote}

That is it. A human language description of the goals and guardrails, along with a description of some APIs to use as tools, becomes a living software system. This is also built on a prompt that anyone can write because they don't have to translate it into source code. 

The other critical ingredient of these living software systems is that they can adapt on the fly. If I hand this travel expense agent a picture of a baseball game, it won't crash with a blue screen, it will just tell me it doesn't know what on earth the picture is. This ability to adapt on the fly builds much more robust systems that are alive and can respond to change and unexpected events -- unlike our current brittle software systems. 

These living software systems, based on prompts, are going to be interdisciplinary creations. The vast majority of software today requires a highly trained software engineer (or team of them) to translate what the domain expert wants into a piece of software. There is no need for this with these living software systems. The Generative AI can directly translate for the domain expert. The living software systems can even be built to help the domain expert write a great prompt! Every level of translation is covered to help the domain expert directly build the system themselves. 

However, to create these living software systems, we are going to have to train our domain experts how to build them. People will need to learn Prompt Engineering~\cite{white2023prompt}. Prompt Engineering is NOT how to make cute word choices to get fun things to happen. Prompt Engineering is about learning to decompose and express goals in ways that allow Generative AI to reason effectively about them. It is about learning how to train our Generative AI, like an intern, to help us. 

\subsection{Prompt Engineering: Yet More User Training to Adapt to the Computing Interface?}

Wait. Don't I teach a class on Prompt Engineering? Aren't you just teaching people to adapt their problems to another interface to computing? This sounds like an incredible level of hypocrisy. 

Yes. We will have to teach people how to work with this new interface, but in a different way. The goal of Prompt Engineering isn't to futz around with the words in the prompt or to cut/paste from a prompt library. The goal of Prompt Engineering is to help people learn how to express their goals in a much richer format than they are used to for consumption by Generative AI. The goal is to help people learn to support their own problem solving skills by realizing that, accurate expression of the problem itself in text, photos, video, and audio is now a critical task in computing.

Further, the expression isn't just the goal, its the rich context surrounding the goal. No one is currently telling their email clients that they are overwhelmed and feeling anxiety due to potential missed messages. No, context isn't something an email client supports. 

Providing the right context to Generative AI is critical to using it effectively. However, everyone has spent decades using software tools that don't support context. Helping users understand that their context matters will take time. Up until ChatGPT, I hadn't told my email client that "I was worried about being misinterpreted" in my message so that it could help me write a less ambiguous statement. 

Our current software tools also don't talk to us -- or at least -- not really. The difference between a Generative AI tool like, Claude, and traditional Internet search, is that when it produces something I don't like, I can't easily say "no, I don't like that and here is why." I can't tell my email client that "no, I am looking for messages from the other Bob that is the CEO." In fact, if you add extra context to your Internet search, like "I am anxious that Bob misunderstood what I said," you are more likely to get advertisements for anxiety medications than the right results. Current software tools just can't handle context.

Generative AI introduces the ability to build context through conversation. Context is critical for translating our goals into actions. So, people are going to have to learn to talk to Generative AI and help iteratively translate goals into actions through a conversation. Yet, almost no one has ever done this for the last several decades, so we shouldn't expect it to be easy or obvious. 

The problem is, our current computing tools can't do any of this. We have spent decades teaching people that they need to have solved the problem completely, before doing anything with the software. Now, we need to people to engage with computing much earlier in the process. People need to learn how they "can choose" to engage with Generative AI to support them. They need to learn how to lead it to support them in accomplishing their goals.

People also aren't used to explaining problem solving strategies and guardrails to their current dead software tools. You don't configure your email client with a statement like "anything that relates to a medical issue with my child or my child's school goes straight to the top of my inbox." We have to first teach people that this is possible and then how to write effective statements / strategies for Generative AI to use. 

Bring an intern into the office and everyone knows how to talk to them. Everyone can have a conversation with the intern to help them improve their work. Still, many people can benefit from training to learn how to lead and communicate with others to help them accomplish goals. If there wasn't a need for this training, we wouldn't have executive coaches, leadership seminars, rising leaders training programs, or any of the myriad other ways that we teach people to improve in this area. 

Effective Prompt Engineering training is a bit like a cross between "AI Leadership" and "effective problem expression and decomposition". Everyone can benefit. Everyone needs it. 

The reason for the training, however, has changed a lot. We aren't training because we have built a tool to accomplish "Goal X" that is inherently flawed in some contexts and we need to help users work around the flaws. We are teaching people to effectively coach a Generative AI or Agentic AI "intern" to help us solve problems. Just like a real human intern, the way we express the problem, the steps we give them to perform, the context we provide (or don't provide), and the guardrails we put in place on the process dramatically impact the outcome. 

\subsection{Can We Trust Generative AI Not to Make Translation Mistakes?}

Everyone gets way too hung up on Generative AI making mistakes. Yes, it does. However, the focus shouldn't be on firing the intern, the focus should be on: 1) giving the intern the right tasks to help with, 2) phrasing the tasks and the outputs in a way that they can be fact checked, and 3) incorporating the work into our existing human processes for testing and verification. 

Every single organization relies on humans that make mistakes all the time. We have robust processes in place to detect mistakes in the most important places. The key with Generative AI is to incorporate its output into the same types of systems. 

The simplest example is the supposed inability of Generative AI to not be able to give references. This misconception is rooted in incorrect use of the tool. Here is a simple prompt that shows the right way to get supporting references out of Generative AI:

\begin{displayquote}
Here are the sources:

---------------

1. ...some source...

2. ...some other source...

Please summarize the sources and provide quotations from each source supporting the summary. Cite the number of the source above in the format [X], where X is the source number.
\end{displayquote}

This simple prompt will cause the Generative AI to produce an answer with a list of sources underneath key statements. To fact check it, you simply trace the sources back to the original list of sources. Does the supposed source exist? Does it say what the Generative AI said? Do you agree with the summary based on the sources it pulled? That is exactly what we should be doing with human work to fact check it.

We have systems in place to fact check work. We have systems in place to test software. We just need to fit these living software systems into a framework that makes them fact checkable, minimizes the impact of errors, and includes human oversight.

\section{Looking to the Future}

Alan Watts once said, "The only way to make sense out of change is to plunge into it, move with it, and join the dance"~\cite{watts1962wisdom}. This wisdom encapsulates the essence of living software systems. Generative AI can help software embrace the dynamic nature of the world and move with it. We can create computing environments that, to use Watts' words, "join the dance" and move with it. These systems will not resist change but will welcome it, adapting fluidly to new contexts and evolving requirements.

We shouldn't underestimate how radical a shift this will be in computing. We also shouldn't expect things to move instantly, as human adoption of the technology will determine how fast it moves. Just because we can build living software systems, doesn't mean that everyone will have the foresight to do it. If we are stuck looking for short-term gains, cutting and pasting silly prompts about "magic", and believing "a summary" is the best thing we an get out of a meeting transcript, then it will take us a long time to reach these living software systems.  

Not every piece of software needs to be a living software system. There are limitations on how fast and repeatably these systems currently perform tasks. These limitations won't make them right for flying a plane, but they will likely help you plan your next trip that includes a flight and text you to suggest an alternative way home when your flight gets canceled. The best part is that we will have the opportunity to teach these living software systems, based on Agentic AI, how to work with us. I will be able to tell it not to bother booking something that isn't a direct flight from Nashville, leaves in the middle of my son's BMX race, or will stick me in a middle seat. 

\bibliographystyle{plain}
\bibliography{bibfile}

\end{document}